\documentclass[aip,jcp,preprint,showkeys,superscriptaddress]{revtex4-1}
 
\usepackage{graphicx,bm,xcolor,microtype,multirow,amscd,amsmath,amssymb,amsfonts,physics,longtable,wrapfig,txfonts,soul}
\usepackage{adjustbox}
\usepackage[normalem]{ulem}

\usepackage[utf8]{inputenc}
\usepackage[T1]{fontenc}
\usepackage{txfonts}
\usepackage{siunitx}

\usepackage[
	colorlinks=true,
    citecolor=blue,
    breaklinks=true
	]{hyperref}
\begin{document}

\title{
  Optical spectra of small silver clusters with the Bethe-Salpeter formalism: a Reassessment 
} 

\author{Xavier Blase}
\affiliation{ Univ. Grenoble Alpes, CNRS, Institut N\'{e}el, F-38042 Grenoble, France }
\email{xavier.blase@neel.cnrs.fr}
  \author{Ivan Duchemin}
\affiliation{Univ. Grenoble Alpes, CEA, IRIG-MEM-L\_Sim, 38054  Grenoble, France}

\date{\today}

\begin{abstract}
 We study the optical absorption spectra of small Ag$_n$ (n=2,4,6,8) clusters using the Bethe-Salpeter equation (BSE) formalism with a Hamiltonian built from $GW$ quasiparticle energies. Calculations are based on an effective core potential  including the 4\textit{sp}   shells in the valence. Non-self-consistent $G_0W_0$ calculations relying on input Kohn-Sham eigenstates generated with semilocal functionals  lead to  very poor BSE absorption spectra, confirming a previous observation.  However, eigenvalue or quasiparticle self-consistent $GW$ calculations dramatically improves the agreement with experiments and TD-DFT spectra obtained with range-separated hybrid functionals.  The importance of the position of the 4\textit{d} occupied band is emphasized.     
\end{abstract}

\keywords{ \textit{Ab initio} many-body theory; $GW$ formalism ; Bethe-Salpeter equation formalism }

\maketitle

\section{Introduction}

The optical properties of noble-metal nanoparticles are dominated by localized surface plasmon resonances (LSPR)  associated   with a strong absorption in the near-UV/visible range, together with very large induced local fields. \cite{Schuller2010} This leads to several applications, including surface enhanced Raman spectroscopy, \cite{Sharma2012} polaritonic chemistry, \cite{Feist2018} enhanced solar cells, \cite{Atwater2010}   biosensing \cite{Anker2008} and cancer therapy. \cite{Bardhan2011} On the theory side, the  analytic treatment of Maxwell equations around a spherical nanoparticle \cite{Mie1908} can be generalized to any shape using e.g. continuum or discrete polarizable models, requiring however an input macroscopic dielectric function $\epsilon(\omega)$ or atomic dynamical polarizabilities $\alpha(\omega)$ usually inferred from the bulk limit (for a recent appraisal of semiclassical approaches, see e.g. Ref.~\citenum{Giovannini2022}).  In the very few nanometers size limit, \textit{ab initio} quantum mechanical approaches become feasible, providing much insight on the atomic structure/properties relation, and further on the interplay between localized \textit{d}-electrons and delocalized \textit{sp} degrees of freedom. Reversely, this delicate balance between   \textit{d} and \textit{sp} contributions to screening and to the optical response represents a severe test case for \textit{ab initio} techniques, from mean-field to more elaborate  but computationally demanding  approaches.

Time-Dependent Density Functional Theory (TD-DFT) \cite{Runge1984,Casida1995} has been one of the tools of choice to study the optical response of silver clusters at the \textit{ab initio} level, \cite{Yabana1999,Zhao2006,Aikens2008,Harb2007,Harb2008,Lecoultre2011,Weissker2011,Bae2012,Anak2014,Rabilloud2013,Rabilloud2014,Koval2015,Weissker2015,Kuisma2015,Gieseking2016,Baseggio2016,Schira2018,Schira2019,Chaudhary2024} due certainly to the excellent accuracy-to-computer-time trade-off.  
It remains nevertheless that the resulting excitation energies may strongly  depend on the choice of the underlying exchange-correlation (XC) functional. Local or semilocal functionals,\cite{Yabana1999,Zhao2006,Idrobo2005,Idrobo2007,Harb2007,Harb2008,Weissker2011,Koval2015,Weissker2015,Baseggio2016} SAOP \cite{Gritsenko1999} or LB98\cite{Baerends1994} functionals with correct asymptotic behavior, \cite{Aikens2008,Bae2012,Kuisma2015,Gieseking2016} and further global \cite{Lecoultre2011} or range-separated hybrid (RSH) \cite{Rabilloud2013,Anak2014,Rabilloud2014,Schira2018,Schira2019} functionals, were shown to lead to different absorption energies. In particular, TD-DFT spectra with (semi)local functionals are characterized by a dramatically larger number of excitations at low energy as compared to calculations with RSH functionals. \cite{Rabilloud2013} In the search for reference data, EOM-CCSD calculations were performed for small Ag$_n$ clusters, \cite{Koutecky1999,Koutecky2001,Rabilloud2013} but changes in Effective Core Potentials (ECP) and basis set sizes complicate the comparison between various techniques. 

The comparison with experiments is made difficult by the effect of the rare gas matrix encapsulating the clusters  during characterization. \cite{Fedrigo1993,Harbich1993,Conus2006,Harb2008,Lecoultre2011,Schira2018}  The overall effect of  the  matrix results from a  competition between long-range polarization  (dielectric) effects, associated with a redshift of the absorption lines,  and a Pauli repulsion (compression) exerted by the matrix on  the cluster that leads to a blue-shift. \cite{Gervais2004,Gervais2005}   Comparing the excitation energies for the dimer in the gas phase or embedded in a neon matrix reveals an overall small blue shift induced by the matrix (see Table~1 Ref.~\citenum{Rabilloud2013}). For larger clusters, through extensive calculations in a QM/QM/PCM framework, where the QM/QM subsystem contains the cluster and a shell of a hundred explicit neon atoms, it was shown that dielectric effects dominate in larger clusters, while compression effects are important in the small size limit. \cite{Schira2018}  In the case of the Ag$_8$ T$_d$ and D$_{2d}$ isomers, the neon matrix induces a small red-shift of the absorption onset, but a small   blue-shift of the main structures at $\sim$4~eV, \cite{Schira2018} indicating on average a compensation of the matrix effects in this size range. Overall, TD-DFT range-separated hybrids (RSH), and in particular the CAM-B3LYP \cite{Yanai2004} and $\omega$B97x \cite{Chai2008} functionals in their original parametrization, were advocated to lead to the  most accurate absorption spectra as compared to available experimental data. \cite{Rabilloud2013,Anak2014,Rabilloud2014,Schira2019} The energy position of the  \textit{d}-band, but also a partial charge-transfer (CT) character in the \textit{d}$\rightarrow$\textit{sp} contributions, \cite{Rabilloud2013,Rabilloud2014} were invoked to explain the success of TD-RSH calculations.   The same conclusions were drawn in the case of gold clusters. \cite{Koppen2012,Anak2014} 

Still within the TD-DFT framework, calculations were performed recently adopting an alternative real-time TD-DFT+U approach. \cite{Chaudhary2024} The onsite U-energy ($U=4$ eV) was designed to improve the agreement with experiment of the frequency-dependent macroscopic dielectric function $\epsilon(\omega)$ for bulk silver. \cite{Avakyan2020} The on-site $U$ contribution  lowers primarily the energy of the localized 4\textit{d} levels, indicating the influence of the 4\textit{d} band on the optical and plasmonic properties.  A good agreement with experiment was demonstrated for the absorption lines of silver clusters at this TD-DFT+U level.\cite{Chaudhary2024} The connection between the local character of the $U$ correction, and the long-range contribution of range-separated hybrids (RSH), is however not straightforward.  

Aside from mean-field techniques and their extensions, a many-body $GW$ \cite{Hedin1965} and Bethe-Salpeter equation \cite{Salpeter1951,Csanak1971,Strinati1982,Strinati1988,Benedict1998,Rohlfing1998,Albrecht1998} (BSE) study of small silver clusters was pioneered by Tiago and coworkers, \cite{Tiago2009}  presenting however large discrepancies with respect to the observed optical spectra. This is a puzzling result since the BSE formalism has been shown through many benchmark calculations on large molecular test sets to yield a good description of singlet excitations with both localized (Frenkel) and charge-transfer (CT)  characters, \cite{Rocca2010,Blase2011,Baumeier2012,Bruneval2015,Jacquemin2015,Jacquemin2017,Gui2018,Nguyen2019,Liu2020,Loos2021,Forster2022,Waide2024,Holzer2025} leaving aside transitions with a large multiple-excitation weight.
The authors suggested that the pseudopotential they use, with the semicore 4\textit{s} and 4\textit{p} shells frozen in the core, was inadequate due to significant overlap with  the 4\textit{d} shell in the valence. Coming in support to this interpretation, the ionization potential   of the silver cation, associated with Ag 4\textit{d} levels, was shown to be significantly improved by including the semi-core shells in the valence. In the real-space-grid formalism adopted, the inclusion of tight 4\textit{sp} electrons in the valence requires very small grid spacing, increasing significantly the cost of such calculations, and the study of Ag$_n$ clusters ($n>1$) with this more accurate pseudopotential was not performed.

In the present study, using 28-core-electrons ECP leaving the 4\textit{sp} degrees of freedom in the valence, and with large Gaussian atomic basis sets, we perform  $GW$ and BSE calculations of the electronic and optical properties of small closed-shell Ag$_n$ silver cluster, with n=2,4,6,8. We confirm the inaccuracy of BSE calculations based on non-self-consistent $G_0W_0$ calculations starting with Kohn-Sham eigenstates generated with (semi)local functionals. However, using  eigenvalue-self-consistent or quasiparticle-self-consistent $GW$ calculations, the resulting BSE spectra come in much better agreement with experiments and with the best TD-DFT results based on the CAM-B3LYP or $\omega$B97x  range-separated hybrid functionals. 
Further, thanks to self-consistency at the $GW$ level, the BSE spectra are very much independent of the starting functional used to generate the needed input one-body eigenstates. Of specific importance for the resulting   absorption spectra, we show  that the energy position of the 4\textit{d}-band is quite sensitive to self-consistency at the $GW$ level. The present results  offer an alternative route to study the optoelectronic and plasmonic properties of larger silver clusters with a weak dependence on the input functional, provided that some level of self-consistency is preserved.

\section{Theory}

We briefly introduce the $GW$ and BSE Green's function many-body perturbation theories (MBPT) emphasizing the similarities and differences with the DFT/TD-DFT formalisms. Thorough derivations can be found in the literature for the $GW$ \cite{Hedin1965,Farid99,Str80,Hybertsen1986,God88,ReiningBook,Ping2013,Golze2019} and BSE \cite{ReiningBook,Salpeter1951,Csanak1971,Strinati1982,Strinati1984,Benedict1998,Rohlfing1998,Albrecht1998,Blase2018,Blase2020} approaches. In a nutshell, the $GW$ formalism is designed to provide well-defined electron energy levels, labeled quasiparticle energies, matching electron addition or removal energies. Further, the BSE formalism tackles the calculation of optical absorption spectra. The electron-hole BSE Hamiltonian is built from the $GW$ quasiparticle energies and the static non-local screened Coulomb potential $W({\bf r},{\bf r}'; \omega=0)$ that replaces the TD-DFT exchange-correlation kernel. The ($-W$) operator serves as a non-local attractive screened interaction between the hole and the electron, allowing to deal with Frenkel, Rydberg or charge-transfer (CT) excitations, without any adjustable parameters to describe short- to long-range electron-hole interactions.

\subsection{The $GW$ and Bethe-Salpeter equation formalisms}

\noindent \textit{Green's function. \;}While density functional theory (DFT) takes the electronic charge density as the central quantity, the $GW$ and BSE formalisms build on  the time-ordered one-body Green's function $G$. Neglecting spin-variables for simplicity, $G$ adopts  the following spectral representation:
\begin{align}
    G({\bf r},{\bf r}'; \omega) = \sum_n \frac{f_n({\bf r}) f_n^{*}({\bf r}') }{ \omega - \varepsilon_n + i \eta \times \text{sign}(\varepsilon_n-\mu) }
\end{align}
where $\mu$ is the chemical potential and $\eta$ a positive infinitesimal. The energies $\lbrace \varepsilon_n \rbrace$ are formally the electron removal energies: $\; \varepsilon_n = E_0[N] - E_n[N-1] \;$ for $\varepsilon_n < \mu$, or the addition energies: $\; \varepsilon_n = E_n[N+1] - E_0[N] \;$ for $\varepsilon_n > \mu$. $E_0[N]$ is the ground-state energy of the N-electron system, while $E_n[N-1]$ and $E_n[N+1]$ are the n-th excited state energies of the (N-1) and (N+1) electron systems, respectively. As such, the $\lbrace \varepsilon_n \rbrace$ can be compared to the  energies measured by direct or inverse   photoemission spectroscopy experiments, providing a good definition of electron energy levels. For (n=0), this gives in particular the lowest ionization potential (IP) and largest electronic affinity (EA) within a minus sign. The $f_n({\bf r})$ are called the Lehmann weights that measure how much the n-th excited state of the (N-1)/(N+1) electron system, with an electron added/removed in ${\bf r}$, overlaps with the N-electron ground state. In the case of mono-determinental many-body wavefunctions, they identify to the standard one-body molecular orbitals (MOs). It can be shown formally that  the $\lbrace \varepsilon_n, f_n \rbrace $ are solutions of a one-body equation:
\begin{align*}
  (\hat{h}_0  + V_H)({\bf r})   f_n({\bf r}) + \int d{\bf r}' \; \Sigma({\bf r},{\bf r}'; \varepsilon_n) f_n({\bf r}) = 
   \varepsilon_n f_n({\bf r})
\end{align*}
where $\hat{h}_0$ is the standard mean-field Hamiltonian containing the kinetic energy and the ionic potential, while $V_H$ is the Hartree potential. The $\Sigma$ operator accounts for exchange and correlation and is called a self-energy. Importantly, it is a non-local and energy-dependent  operator. Namely, in contrast with the DFT Kohn-Sham (KS) equation, states with different energies do not see the same XC potential. As such, the self-energy is state-specific. \\


\noindent \textit{The $GW$ approximation. \;} The $GW$ formalism \cite{Hedin1965} proceeds by merging Green's function and linear-response theories, showing that $\Sigma$ can be related to the variation $\partial G(12)/\partial U(3)$ where $U$ is an external local perturbing potential. This connection leads to the  self-consistent Hedin's equations  \cite{Hedin1965} that, \textit{to the lowest order in the screened Coulomb potential W}, yields the $GW$ self-energy:
\begin{align*}
    \Sigma^{GW}({\bf r},{\bf r}';E) = i \int \frac{d\omega}{2\pi} \; e^{-i \eta \omega} G({\bf r},{\bf r}';\omega+E) W({\bf r},{\bf r}';\omega )
\end{align*}
where $\; W =  v + v \chi_0 W \; $ is the screened Coulomb potential in the random-phase approximation (RPA), with $v$ the bare Coulomb potential. The operator $\chi_0$ is the irreducible electronic susceptibility. In a nutshell, the link with linear response theory allows in particular to account for the electronic relaxation induced by adding/removing an electron to the system. 

A standard $GW$ calculation starts by building $G$ and $\chi_0$ from an available Hartree-Fock or Kohn-Sham (KS) calculation, reducing the Lehmann weights and addition/removal energies to the Kohn-Sham (KS) molecular orbitals (MOs) and eigenvalues, with for the susceptibility:
\begin{align*}
    \chi_0({\bf r},{\bf r}'; \omega) &= \sum_{mn} \frac{(f_m-f_n) \; \phi_n^{KS}({\bf r}) \phi_m^{KS}({\bf r}) \phi_m^{KS}({\bf r}') \phi_n^{KS}({\bf r}') } { \omega - (\varepsilon_n^{KS} - \varepsilon_m^{KS}) + i\eta \times \text{sgn}( \varepsilon_n^{KS} - \varepsilon_m^{KS}) }
\end{align*}
 We keep here the MOs real for finite size systems. The $(f_n,f_m)$ are occupation factors. This defines the single-shot non-self-consistent $G_0W_0$@XCF self-energy-operator, where @XCF indicates which XC functional has been used to generate the input Kohn-Sham eigenstates. The $G_0W_0$@XCF quasiparticle energies $\lbrace \varepsilon_n^{QP} \rbrace$ can be simply obtained by replacing in the input Kohn-Sham energy levels $\lbrace \varepsilon_n^{KS} \rbrace$ the DFT XC potential $V^{XC}_{DFT}$ contribution by that of the $\Sigma=iG_0W_0$ self-energy:
 \begin{align}
     \varepsilon_n^{QP} = \varepsilon_n^{KS} + \langle \phi_n^{KS} | \Sigma( \varepsilon_n^{QP} ) -V^{XC}_{DFT} | \phi_n^{KS} \rangle 
 \end{align}
 This approach significantly improves the adequation of the $\lbrace \varepsilon_n^{QP} \rbrace$   quasiparticle energies   to true electron addition/removal energies, but leaves some dependence on the chosen starting XC functional. Partial self-consistency can be obtained by reinjecting self-consistently the corrected energy levels in the construction of the $G$ and $\chi_0$ operators.\cite{Hybertsen1986,Blase2011,Shishkin2007,Kaplan2016,Rangel2016} This scheme will be labeled ev$GW$@XCF in the following. The dependence on the starting  XC functional is significantly removed, but no completely since the Kohn-Sham molecular orbital $\lbrace \phi_n^{KS} \rbrace$ are kept frozen.  Rediagonalizing the full $\; \mathcal{H^{QP}} = h_0 + V_H + \Sigma \;$ operator in the Kohn-Sham MO basis allows  updating as well the one-body MOs, leading to a quasiparticle-self-consistent qs$GW$ scheme. \cite{Kotani2007,Marie2023,Duchemin2025} As a variation on the historical Kotani, Van Schilfgaarde, Faleev qs$GW$ scheme, \cite{Kotani2007} we will be using the recently introduced  Joint Approximate Diagonalization (JAD) qs$GW$    approach\cite{Duchemin2025}  that does not require reducing the self-energy to a symmetrized static Hermitian form. The qs$GW$ scheme leads to results strictly independent of the starting  XC  functional. 

Adopting a Coulomb-fitting resolution-of-the-identity and analytic continuation  techniques (see Technical details below), $G_0W_0$ and ev$GW$ calculations  scales as $\mathcal{O}(N^4)$.  Within the same framework, quasiparticle self-consistent qs$GW$ calculations are more expansive, scaling as $\mathcal{O}(N^5)$. \\
 
\noindent  \textit{BSE formalism. \;} While TD-DFT hinges on the electronic susceptibility $\chi$ as the derivative of the charge density $\rho$ by an external local potential $U$, the Bethe-Salpeter equation (BSE) formalism defines a generalized 4-point response function $L$ as the derivative of the one-body time-ordered Green's function by a non-local perturbation, namely: 
\begin{equation}
    \chi(1,2) \stackrel{DFT}{=} \frac{ \partial \rho(1) }{\partial U(2)} 
    \longrightarrow 
    L(1,2;1',2') \stackrel{BSE}{=} \frac{ \partial G(1,1') }{\partial U(2',2)} 
\end{equation}
We adopt here space-time notations with e.g. $1=({\bf r}_1 t_1)$. 
Replacing further the DFT XC potential by the $GW$ self-energy leads to a modified XC kernel associated with
the derivative $\partial GW / \partial G$. Expressed in the standard $\phi_i\phi_a$ transition basis, where the $(i/a)$ indices point to occupied/virtual MOs, one obtains a set of equations very similar to that of the Casida's formulation for TD-DFT,\cite{Casida1995} that is:
\begin{align}
    \begin{pmatrix}
        R & C \\
        -C^{*} &-R^{*} 
    \end{pmatrix}
    \begin{pmatrix}
        X^{\lambda} \\
        Y^{\lambda} 
    \end{pmatrix}
    = \Omega_{\lambda}
    \begin{pmatrix}
        X^{\lambda} \\
        Y^{\lambda} 
    \end{pmatrix}
\end{align}
with  two-body electron-hole eigenstates:
\begin{align}
    \psi^{eh}_{\lambda}( {\bf r}_e, {\bf r}_h) = \sum_{ia} \left[ X^{\lambda}_{ia} \phi_i({\bf r}_h) \phi_a({\bf r}_e) + Y^{\lambda}_{ia} \phi_i({\bf r}_e) \phi_a({\bf r}_h) \right]
\end{align}
In a standard BSE calculation, restricting the screened Coulomb potential to its static limit (an adiabatic approximation)  \cite{beyondadiabatic} and neglecting the higher-order $( \partial W / \partial G)$ term, \cite{higherorder} the diagonal and off-diagonal blocks read:
\begin{align}
    R_{ia,jb} &= (\varepsilon_a^{GW} - \varepsilon_i^{GW}) \delta_{ia} \delta_{jb} + \kappa (ia|jb) - W_{ij,ab} \\
    C_{ia,jb} &=   \kappa (ia|bj) - W_{ib,aj}
\end{align}
with $\kappa=2$ for singlets and:
\begin{align}
    W_{ij,ab} = \int d{\bf r} d{\bf r}' \; \phi_i({\bf r}) \phi_j({\bf r})  W({\bf r},{\bf r}'; \omega=0) 
    \phi_a({\bf r}') \phi_b({\bf r}')
\end{align}
The BSE formalism in its adiabatic formulation is thus very  similar to TD-DFT but with $GW$ quasiparticle energies in place of Kohn-Sham eigenvalues, and with the non-local screened Coulomb-potential $W$ instead of the TD-DFT XC kernel. As such, the BSE formalism and TD-DFT are associated with similar costs once the $GW$ calculation is performed. 
In the following, we adopt the notation BSE/$G_0W_0$@XCF, BSE/ev$GW$@XCF or BSE/qs$GW$ to indicate from which $GW$ calculation the BSE Hamiltonian is built.

\subsection{Technical details}

The many-body $GW$ and BSE calculations are performed with the {\sc{beDeft}} (beyondDFT) code,
\cite{Duchemin2020,Duchemin2021} adopting Gaussian basis sets within a Coulomb-fitting (RI-V) resolution-of-the-identity \cite{Vahtras1993,Ren2012,Duchemin2017} approach. The energy integral defining the dynamical $GW$ self-energy is performed using a scheme that combines the contour deformation approach with an analytic continuation of the screened Coulomb potential. \cite{Duchemin2020} Such an approach was shown to be preferable over continuing directly the self-energy that is much more structured. The screened Coulomb potential $W(z)$ is calculated for a set of 12 optimized frequencies along the imaginary axis. Since the quasiparticle self-consistent qs$GW$ approach mixes states located far away from the imaginary-frequency axis, the frequency sampling set is completed by a  ($z_n=n \Delta E +i\eta$) frequency-grid located typically $\eta\simeq 1.5 \times \Delta E$ above the real axis, choosing $\Delta E=2$ eV.\cite{Duchemin2020} Increasing the density of these sampling points, or comparing with the more expansive standard contour-deformation approach,  allows ascertaining that meV accuracy can be obtained with this technique (see also Refs.~\citenum{Golze2023,Kehry2023} for the validation of this approach in the case of core-levels).

The quasiparticle self-consistent $GW$ calculations (qs$GW$) are performed using the recently developed Joint Approximate  Diagonalization (JAD) scheme \cite{Duchemin2025} allowing to perform qs$GW$ calculations without restricting the self-energy to a static symmetrized self-energy as originally introduced by  Kotani,   van Schilfgaarde and Faleev. \cite{Kotani2007} Both approaches were however shown to agree within a mean-absolute difference of 60~meV for the ionization potential (IP) over a large organic molecular test set.\cite{Duchemin2025} These qs$GW$ calculations are compared to eigenvalue-self-consistent (ev$GW$) calculations. \cite{Hybertsen1986,Blase2011,Shishkin2007,Kaplan2016,Rangel2016}  
In what follows, our ev$GW$@XCF calculations  are performed using the family of hybrid functionals mixing the PBE  XC  functional \cite{Perdew1996} with exact exchange. Besides PBE, this includes the PBE0 functional, \cite{Adamo1999,Ernzerhof1999} and the system-dependent PBEh($\alpha$) functional where the amount $\alpha$ of exact exchange is tuned so that the Kohn-Sham (KS)  and resulting $GW$ gaps are equal. We perform as well non-self-consistent $G_0W_0$@PBE calculations to analyze the difficulties revealed in Ref.~\citenum{Tiago2009} when performing BSE calculations on top of $G_0W_0$@LDA calculations. 

Our BSE calculations are very standard, adopting the adiabatic limit that restricts the BSE kernel to the  low-frequency limit of the screened  Coulomb potential,   neglecting further the higher-order $(\partial W / \partial G)$ derivative in the $( \partial  \Sigma^{GW} / \partial G)$ kernel contribution.  We go beyond the Tamm-Dancoff approximation, keeping the non-resonant components of the BSE electron-hole eigenstates. All occupied (including core states) and unoccupied molecular orbitals are used when constructing the two-body BSE Hamiltonian.  

Input Kohn-Sham eigenstates, needed for $GW$ and BSE calculations, together with TD-DFT calculations, are generated with the Orca package. \cite{orca} TD-DFT calculations on silver clusters have been performed by several groups in the past and we just reproduce these results for sake of comparison with BSE. We compare in particular BSE data to TD-DFT calculations performed with the PBE functional\cite{Perdew1996} and with two range-separated hybrids (RSH), the CAM-B3LYP \cite{Yanai2004} and $\omega$B97X \cite{Chai2008}  functionals. These two later functionals were shown to produce the best agreement with experiment within the TD-DFT framework. \cite{Rabilloud2013,Rabilloud2014} 

\begin{figure}[t]
	\includegraphics[width=8.6cm]{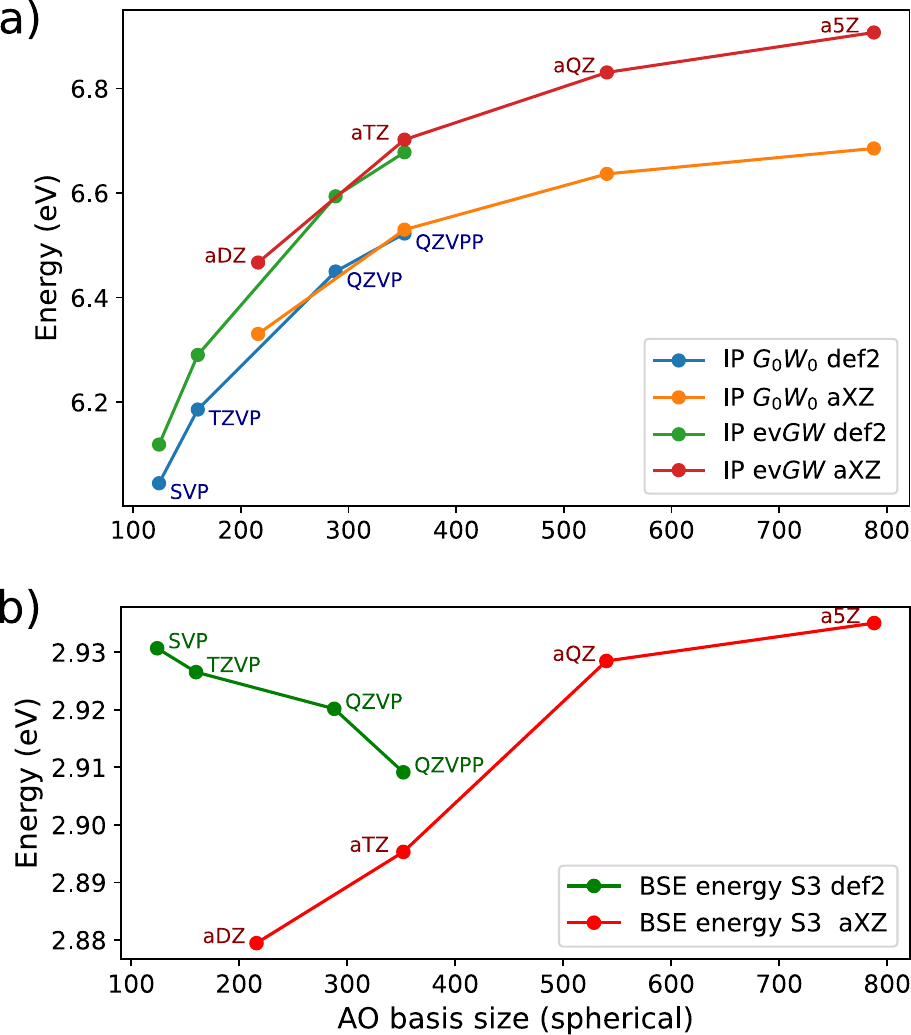} 
	\caption{ Convergence rates as a function of the size of the basis set in the def2 and \emph{aug}-cc-pVXZ (aXZ) families for the Ag$_4$ cluster.
              (a) Ionization potential at the $G_0W_0$@PBE and ev$GW$@PBE levels.
              (b) Lowest bright absorption line (S3) at the BSE/ev$GW$@PBE level. 
              The size of the basis sets are given in their spherical representation. 
              The difference of energy between the a4Z and a5Z data amounts to 76~meV and 7~meV for the ev$GW$@PBE IP and the BSE/ev$GW$@PBE (S3) excitation energy, respectively.}
	\label{fig:fig1}
\end{figure} 

$GW$ and BSE calculations are explicitly correlated methods showing a slower convergence rate with basis set size as compared to (TD-)DFT. We show in Fig.~\ref{fig:fig1}(a) the evolution of the $GW$ ionization potential IP (negative of the HOMO energy) as a function of basis set size in the ``def2"   \cite{Weigend2005a} and \emph{aug}-cc-pVXZ-PP (aXZ) \cite{Peterson2005} families. Both are associated with  Effective Core Potentials (ECP) containing 28 frozen electrons, \cite{Andrae1990,Figgen2005} leaving the 4\textit{sp} degrees of freedom in the valence. This is an important issue to obtain proper 4\textit{d} levels.\cite{Tiago2009} In the chosen RI-V scheme, the def2 and aXZ basis sets are associated with their corresponding optimized auxiliary basis (e.g. def2-TZVP-RI \cite{Hellweg2007} and aug-cc-pVQZ-PP-OPTRI \cite{Figgen2005}) as available from the Basis Set Exchange website. \cite{Pritchard2019}  At the $GW$ level, the def2 and aXZ IP energy data show very consistent results at equivalent basis set size. We adopt in the following the aQZ basis set that allows getting $GW$ quasiparticle energies within 0.1~eV of the complete basis set limit (see Caption Fig.~\ref{fig:fig1}). We provide in the Supplementary Material (SM)   an analysis of the evolution of the ev$GW$@PBE and qs$GW$  IP and EA energies from the aTZ to the aQZ basis set, together with the evolution of the onset of absorption at the BSE/qs$GW$ level. The qs$GW$ scheme is shown to converge significantly faster than ev$GW$@PBE, a behavior that we tentatively assign to the variational nature of qs$GW$.\cite{Beigi2017}  
In particular, aTZ qs$GW$ IP / EA values are within $\sim$70~meV/40~meV of the aQZ values. 
Concerning the BSE/qs$GW$ data, the aTZ values are within a very few meVs of the aQZ data, indicating that BSE/qs$GW$ data are converged at the aTZ value. As such, we will restrict our BSE/qs$GW$ calculations to the aTZ basis set for the largest (n=8) clusters. Overall, the convergence is faster at the BSE level due to a similar evolution with basis set size of the occupied and virtual energy levels.

The cluster geometries are the B3LYP/aDZ ones as provided in Ref.~\citenum{Chen2013}.
We limit our study to close-shell clusters with an even number of atoms. The Ag$_4$ and Ag$_6$ clusters are the D$_{2h}$   and D$_{3h}$  planar isomers, respectively. Concerning the   Ag$_8$ cluster, we study the T$_d$ (8a) and D$_{2d}$ (8b) isomers, the $T_d$ isomer being the lowest energy at the B3LYP/aDZ level, but the D$_{2d}$ isomer  optical absorption spectrum resembling more closely the experimental one at the TD-CAM-B3LYP or TD-$\omega$B97x level.\cite{Rabilloud2013}
We emphasize that the B3LYP/aDZ functional tends to slightly overestimate the Ag bond length. In the dimer case,
the B3LYP/aDZ bond length amounts to 2.5850~\AA~that can be compared to 2.5335~\AA~experimental value. Such a bond length overestimation  leads to a slight red-shift of the excitation energies, from 2.95~eV to 2.91~eV for the absorption onset (BSE/qs$GW$ aQZ values). 


\section{Results}

\subsection{Quasiparticle energies}

We present in Table~\ref{tab:table1} the aQZ $GW$ ionization potential (IP) and electronic affinity (EA) defined as the opposite of the $GW$ quasiparticle HOMO and LUMO energies. Experimental values are from Refs.\citenum{Jackschath1992,Ho1990}. 
The analysis of the IP values for the small Ag$_n$ clusters (n=2,4,6,8a,8b) hardly allows to discriminate between the non-self-consistent and self-consistent schemes.
On the average, the $G_0W_0$@PBE   and self-consistent  $GW$ schemes yield rather similar results in good agreement with the available experiments. For example, the mean-absolute-deviation (MAD) between $G_0W_0$@PBE   and   qs$GW$ amounts to 0.09~eV, while $G_0W_0$@PBE and qs$GW$ agree with experiments within 0.13~eV and 0.16~eV MAD values. A full map of the MAD and mean-signed deviations (MSD) between methods is given in the SM.  The present aQZ values are estimated to be within 0.1-0.2 eV below the CBS limit, but the experimental values are also deemed to be upper-bounds for the vertical IP.\cite{Jackschath1992,Tiago2009} As such, in the absence of quantum chemistry reference values in the large basis set limit and performed on the very same geometries, one can only conclude that the  agreement with experiment is satisfactory. 

\begin{table}
\begin{center} 
\begin{tabular}{ c|c|ccccc|c} 
\hline
                 \multicolumn{8}{c}{ Ionization potential (eV)}    \\
\hline
   n    & Exp.  & $G_0W_0$@ & \multicolumn{3}{c}{ev$GW$@} &  qs$GW$  & $\Delta$SCF  \\
        &       & PBE      &  PBE     &  PBE0   &  PBEh50 &          &    B3LYP    \\
 2      & 7.60  & 7.79     &  8.06    &   7.78  &  7.57   &  7.68    &  7.85            \\
    4   & 6.65  & 6.65     &  6.84    &   6.60  &  6.39   &  6.53    &  6.50        \\
    6   & 7.15  & 7.13     &  7.58    &   7.27  &  7.03   &  7.24    &  7.06        \\
    8a  & 7.10  & 6.99     &  7.41    &   7.04  &  6.76   &  [6.98]  &  6.83        \\
    8b  &       & 6.79     &  7.14    &   6.76  &  6.49   &  [6.71]  &  6.60     \\   
\hline
               \multicolumn{8}{c}{ Electronic Affinity (eV)}    \\
\hline
    2   &  1.06 &  1.36    &   1.02   &   0.93  &  0.85   &  0.88    &  0.99         \\
    4   &  1.65 &  2.08    &   1.69   &   1.50  &  1.33   &  1.42    &  1.55  \\
    6   &  2.06 & 1.80     &   1.44   &   1.23  &  1.05   &  1.15    &  1.24  \\
  8a    &  1.85 &  1.65    &   1.31   &   1.08  &  0.92   &  [0.99]  &  1.07 \\
  8b    &       &  1.91    &   1.62   &   1.43  &  1.26   &  [1.34]  &  1.29    \\
 \hline
               \multicolumn{8}{c}{ Photoemission gap (IP-EA) (eV) }    \\
 \hline

    2   & 6.54  & 6.43     &   7.04   &   6.85  &  6.71   &  6.81     & 6.86   \\
    4   & 5.00  & 4.55     &    5.15  &   5.09  &  5.05   &  5.11     & 4.95  \\
    6   & 5.09  & 5.33     &   6.14   &   6.03  &  5.98   &  6.08     & 5.82  \\
  8a    & 5.25  & 5.34     &   6.10   &   5.95  &  5.85   &  [5.99]   & 5.77  \\
  8b    &       & 4.88     &  5.52    &   5.33  &  5.23   &  [5.37]   &  5.31    \\
\end{tabular}
\end{center}
\caption{ Ionization potential, electronic affinity, and resulting photoemission gap, at various $GW$ levels  (aQZ basis set), together with available experimental data. 
The  Ag$_8$ 8a/8b isomers are in the T$_d$/D$_{2d}$ geometry. Numbers in brackets are the  aQZ  estimated data obtained by adding 0.07 eV/0.04 eV to the aTZ IP/EA results (see analysis in the SM). Experimental data from Ref.~\citenum{Ho1990} for the EA and Ref.~\citenum{Jackschath1992} for the IP. 
We also report $\Delta$SCF B3LYP (aTZ) calculations.
} 
\label{tab:table1}
\end{table}

 \begin{figure}[t]
\includegraphics[width=8.0cm]{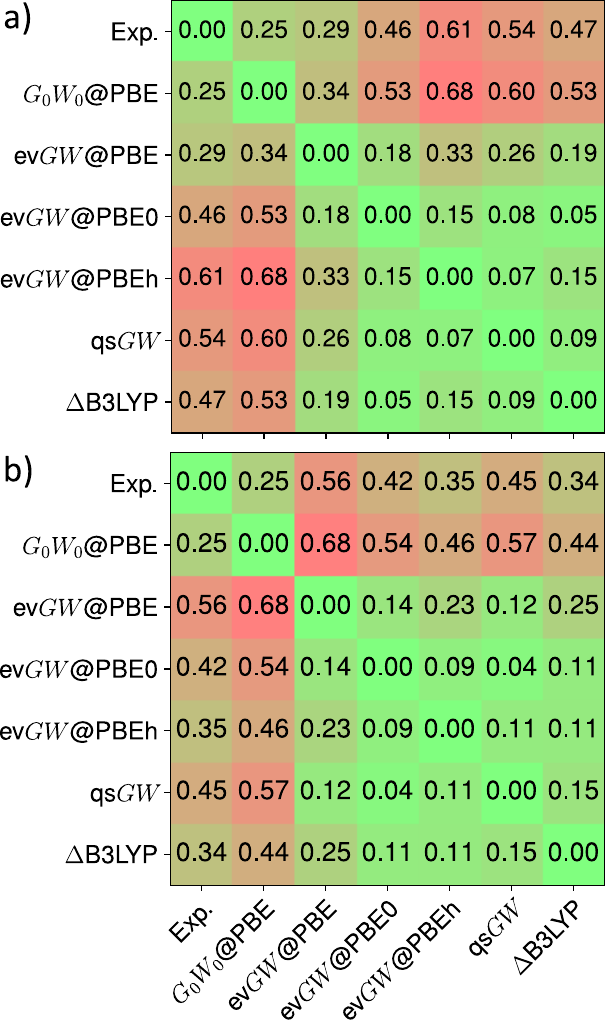}
	\caption{Matrix representation of the mean-absolute-deviation (MAD) between methods for (a) the electronic affinity (EA)  and (b)  the (IP-EA) gap for the Ag$_n$ clusters (n=2,4,6,8a,8b).
    Exp. points to the available experimental data.
    The PBEh functional is PBEh(0.5) with 50$\%$ of exact exchange.  
    $\Delta$B3LYP indicates the $\Delta$SCF B3LYP (aTZ) approach. 
    The MAD values (black numbers) are given in eV. The background color goes from green, representing small deviations, to red for the larger deviations. }
	\label{fig:fig2}
\end{figure}

Concerning now the EA, the non-self-consistent $G_0W_0$@PBE scheme yields values significantly larger than the one obtained at the self-consistent levels (see Table~\ref{tab:table1} and MAD Fig.~\ref{fig:fig2}a). As a result, $G_0W_0$@PBE is on the average in  better agreement with the available experimental data. In contrast, all self-consistent approaches are in good agreement with each other and with a standard $\Delta$SCF B3LYP (aTZ) approach. 
 We note that the larger MADs between available experimental data and the self-consistent schemes or $\Delta$B3LYP originate  mainly from the Ag$_6$  and   Ag$_8$ clusters. For the smaller Ag$_2$ and Ag$_4$ clusters, the self-consistent calculations are actually closer to available experimental values. Concerning these later, they stem  from a photoelectron spectroscopy of the anions, \cite{Ho1990} and the comparison between theory and experiment relies on the assumption that neutral and anionic clusters have reasonably close geometries. Concerning e.g. the Ag$_8$ cluster, we find that the T$_d$ isomer, the most stable neutral ground-state structure, becomes less stable than the D$_{2d}$ isomer in the relaxed anionic form,  at least at the  B3LYP/aDZ level (see anionic energies and geometries in the SM). The extensive search for the lowest energy anionic structures is beyond the scope of the present paper. 
 In the absence again of  quantum chemistry data for the same isomers,  geometries, basis sets and ECP, one can conclude that the agreement with experiment is fair even though rather scattered.

 We provide for information the resulting data for the (IP-EA) photoemission gaps. As a consequence mainly of the EA values, the $G_0W_0$@PBE gaps are closer on the average to the available experimental data as compared to self-consistent calculations. The related MADs are given in Fig.~\ref{fig:fig2}b. As compared to qs$GW$, the non-self-consistent $G_0W_0$@PBE scheme yields gaps smaller by -0.57~eV  (MSD value, see SM). Here again, self-consistent $GW$ schemes and the $\Delta$SCF B3LYP approach yields very similar gaps. Since the difference between virtual and occupied energy levels enters directly the BSE Hamiltonian, we will come back to these data when discussing BSE absorption spectra.

\begin{figure}[b]
\includegraphics[width=8.0cm]{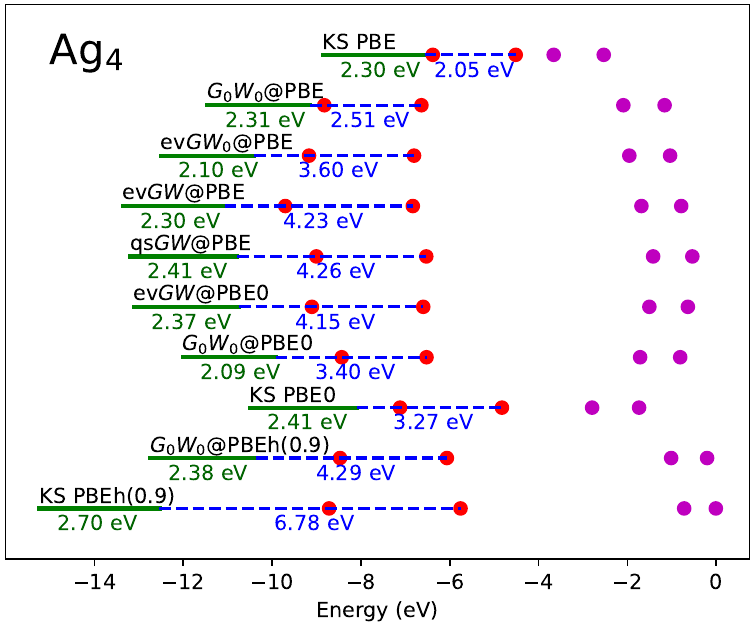}
	\caption{  Schematic representation of the energy position of the \textit{d}-band (green line), (HOMO-1) and HOMO states (red dots) and LUMO and (LUMO+1) states (cyan dots) for Ag$_4$ at different levels of approximation (aQZ basis set). The dashed blue line and blue number indicate the energy difference between the top of the occupied \textit{d}-band and the HOMO level. The green line and number indicate  the \textit{d}-band width. }
	\label{fig:fig3}
\end{figure}

Besides the position of the 5\textit{sp}-like states controlling the gap, we now consider the energy position of the 4\textit{d} occupied band, focusing in Fig.~\ref{fig:fig3} on the Ag$_4$ cluster.   The  (HOMO-1) and HOMO MOs are indicated by red dots, while the cyan dots locate the position of the LUMO and (LUMO+1) MOs.  The 4\textit{d}-band and its width are represented in green. Taking the self-consistent $GW$ data as a reference, the 4\textit{d}-band is located at too high an energy at the KS PBE level. Further, the energy difference (blue numbers)  between the 4\textit{d}-band and the HOMO is dramatically too small at the KS PBE level. The $GW$ correction to the KS eigenvalues is thus much larger for the 4\textit{d}-band as compared to the \textit{sp}-related HOMO and LUMO levels.  
Such a destabilization of the KS PBE localized 4\textit{d} MOs can be related to the strong self-interaction error induced by (semi)local functionals, an error reduced at the $GW$ level as already documented in the bulk limit. \cite{Marini2002} These results are also consistent with the DFT study of the silver Ag(111) surface using an orbital-dependent exchange-correlation functional beyond LDA, \cite{Thygesen2011} showing that semilocal functionals locate the 4\textit{d}-band too high in energy, overscreening the \textit{sp} surface plasmon. Likewise, the DFT+U formalism proposed in Refs.~\citenum{Avakyan2020,Chaudhary2024} allowed stabilizing selectively the localized 4\textit{d}-band. 

Scrutinizing now the various $GW$ calculations, a salient feature of the non-self-consistent $G_0W_0$@PBE calculation is that the energy difference between the top of the 4\textit{d}-band and the HOMO level remains severely underestimated as compared to self-consistent $GW$ data. This energy difference ranges from 2.05 eV, 2.51 eV, and 4.23 eV, at the KS PBE, $G_0W_0$@PBE and ev$GW$@PBE levels. In absolute value with respect to the vacuum level, the ``single-shot" $G_0W_0$@PBE calculation stabilizes the 4\textit{d}-band as compared to KS PBE data, but not sufficiently as compared to self-consistent $GW$ calculations. Namely, the position of the 4\textit{d}-band much strongly discriminates between self-consistent schemes and non-self-consistent calculations starting with (semi)local XC functionals at the DFT level.   

The partially self-consistent ev$GW$@PBE data are close to the qs$GW$ ones, indicating that it is not the update of the shape of the 4\textit{d} MOs that is involved. Freezing (not updating) the screened Coulomb potential, an intermediate scheme labeled ev$GW_0$, we obtain an intermediate description of the 4\textit{d}-band, falling  in between  the  $G_0W_0$@PBE and ev$GW$@PBE  results. This suggests that the reduced participation to screening of the 4\textit{d} orbitals, when pushed to lower energy, participates  self-consistently to its stabilization. We note that this affects much less the \textit{sp} levels (e.g. HOMO and LUMO) illustrating that the screened Coulomb potential is non-local and acts differently on different orbitals. The PBE (LUMO-1) level, that presents an hybrid \textit{sp} and \textit{d}-character, is characterized by an intermediate behavior, resulting in a decoupling with the \text{d}-band in the self-consistent $GW$ cases. We do not observe a strong evolution of the occupied \textit{d}-band  width. We provide in the SM (Fig.~S2) a similar analysis for the largest Ag$_8$ T$_d$ cluster. 

To conclude this exploration, we further start with an optimally-tuned PBEh functional, with 90$\%$ of exact exchange so that the KS gap and the $G_0W_0$@PBEh(0.9) gap are nearly identical for Ag$_4$. Non-self-consistent $G_0W_0$ calculations starting from an optimally tuned functional has been shown to be a very valuable pathway in the study of the IP and EA energies of molecular systems. \cite{Bruneval2013,Rangel2016} We observe however that the \textit{d}-band lands significantly too low in energy as compared e.g. to qs$GW$ data. This illustrates the difficulty of tuning with a single parameter both the HOMO-LUMO gap (of \textit{sp} character) and the position of the \textit{d}-band. A single-shot $G_0W_0$@PBEh(0.9) allows however to put the HOMO-LUMO gap and the \textit{d}-band in better agreement with the qs$GW$ scheme, even though the IP reduces to 6.06 eV, that is $\sim$0.5 eV smaller than  qs$GW$.

\subsection{ Bethe-Salpeter calculations }

\begin{figure}[t]
	\includegraphics[width=0.75\textwidth]{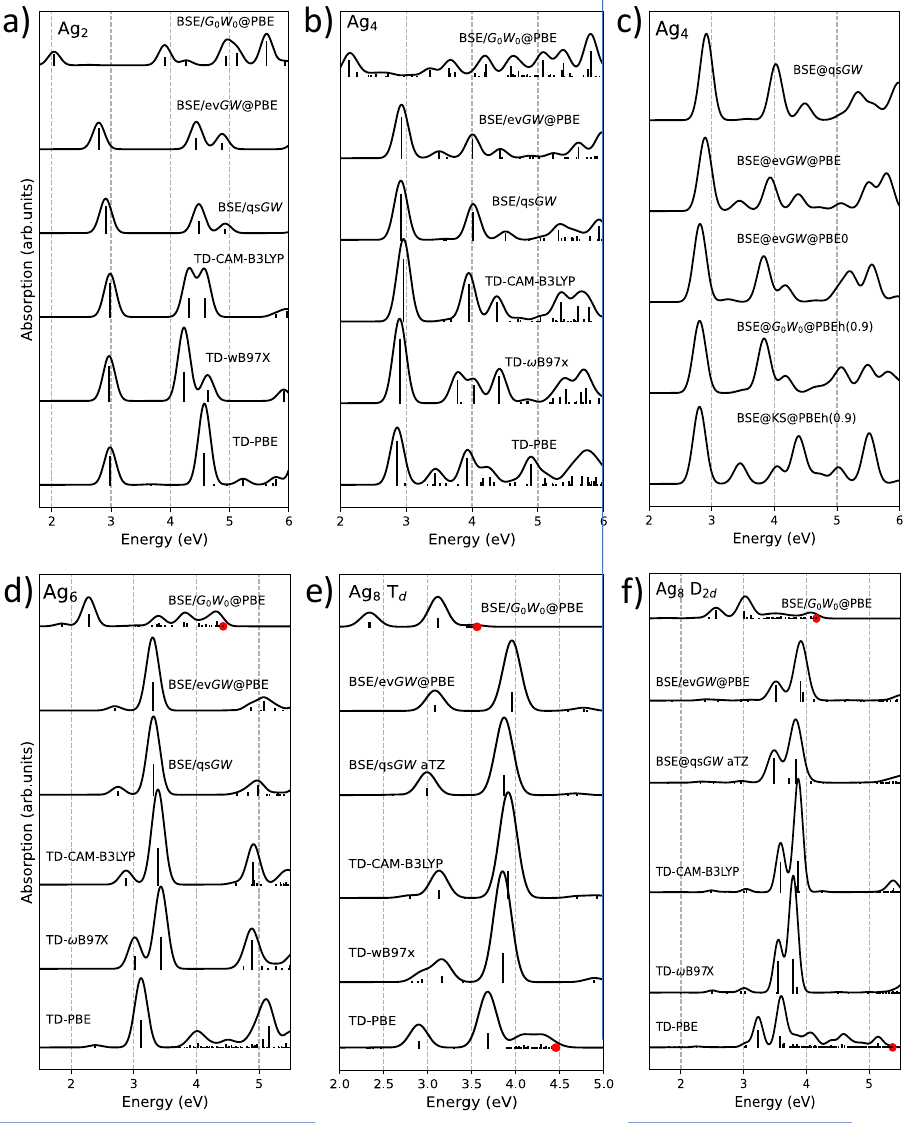}
	\caption{ Absorption spectra for Ag$_n$(n=2,4,6,8a,8b) at the  BSE/$G_0W_0$@PBE, BSE/ev$GW$@PBE, and BSE/qs$GW$ levels (aQZ basis set, except for n=8 BSE/qs$GW$ cases for which the aTZ is used). In (c), we compare for Ag$_4$ the BSE/ev$GW$@XCF spectra with XCF=PBE, PBE0, and the gap-tuned PBEh(0.9), to BSE/qs$GW$ and BSE/PBEh(0.9)
    (aTZ level). 
    These calculations are completed by TD-CAM-B3LYP,  TD-$\omega$B97X and TD-PBE calculations (aTZ basis set). The continuous curves are obtained from the energy and oscillator strengths using a 0.1~eV gaussian broadening. Red dots indicate the highest excitation calculated.  }
	\label{fig:fig4}
\end{figure}

We now explore the optical absorption spectra for which much larger sets of experiments and calculations are available. 
We provide in   Fig.~\ref{fig:fig4} the BSE absorption spectra obtained for Ag$_n$ (n=2,4,6,8a,8b) following   a $G_0W_0$@PBE, an ev$GW$@PBE or a qs$GW$ calculation. 
 We adopt the aQZ basis set for good convergence, except for the 8a and 8b BSE/qs$GW$ data treated at the aTZ level as discussed above.  The results are compared to TD-DFT calculations with range-separated-hybrid functionals, namely the CAM-B3LYP and $\omega$B97X functionals, that have been shown in Refs.~\citenum{Rabilloud2013} to yield the best agreement with experiment comparing various functionals. For sake of illustration, the results of  TD-PBE calculations are also provided. Due to faster convergence with basis set size, the TD-DFT calculations are performed at the aTZ level. We provide  in the SM (Figs.~S4-S5)  a comparison with extracted experimental absorption plots from Ref.~\citenum{Lecoultre2011}. 

Clearly, even though the $G_0W_0$@PBE scheme leads to seemingly reasonable HOMO and LUMO energies as compared to available experimental evidences, the resulting BSE absorption spectra stand  as  severe outliers. Not only the position of the onset of energy is significantly red-shifted, but the overall shape is difficult to reconcile with reference data. This is similar to what was obtained at the BSE/$G_0W_0$@LDA level by Tiago and coworkers \cite{Tiago2009} who concluded in a ``poor agreement with experiment". This former exploration of the optical properties of silver clusters at the many-body level relied on a pseudopotential incorporating the 4\textit{sp} semicore shells in the frozen core, but the same conclusion applies here when a more reliable ECP is used. 

We observe that the $G_0W_0$@PBE gaps (Table~\ref{tab:table1}) are found to be either larger or smaller than the available experimental values, showing thus no correlation with the systematic redshift of the BSE spectra. A better rational can be found by noticing that  the $G_0W_0$@PBE gaps are systematically smaller than their qs$GW$  analogs as discussed above (MSD -0.57~eV, see SM). However, the overall shape cannot be reconciled either with BSE/qs$GW$ or TD-CAM-B3LYP. The erroneous position of the 4\textit{d}-band at the $G_0W_0$@PBE level can certainly explain that the discrepancy goes beyond a simple rigid shift.  

Following further the analysis provided in Ref.~\citenum{Rabilloud2013}, taking Ag$_4$ as an example, we  find that the BSE/$G_0W_0$@PBE spectrum accumulates 133 states below 6~eV, that can be compared to the 46, 44, 53 and 51 transitions in the same energy range for the BSE/evGW, BSE/qs$GW$, TD-CAM-B3LYP and TD-$\omega$B97x levels. 
As such, the BSE/$G_0W_0$@PBE spectrum is associated with a collapse of a large number of transitions at low energy.
This is observed for all clusters.  Turning e.g. to Ag$_8$ D$_{2d}$ (8b), the BSE/qs$GW$ and TD-CAM-B3LYP spectra contain 66 and 72 excitations below 5.5~eV, while the BSE/$G_0W_0$@PBE locates its 120-th transition at only 4.16~eV only (see red dot, Fig.~\ref{fig:fig4}f). We did not seek to calculate  transitions at higher energy. 

The  BSE/ev$GW$@PBE and BSE/qs$GW$ spectra provide  on the contrary a very good agreement with the TD-DFT data using RSH functionals. We provide in Table S2 of the SM a list of the main absorption lines energy and oscillator strength up to $\sim$5~eV  (the selected transitions are marked by  blue/green dots in Fig.~S4-S5). From the  12 transitions listed for n=(2,4,6,8a,8b) at the theoretical level,\cite{degenerate} for which the identification is unambiguous, the mean-signed-deviation (MSD) and mean-absolute-deviation (MAD) between BSE/ev$GW$@PBE and BSE/qs$GW$ (taking the later as the reference) amount to  7~meV and 55~meV, respectively. This indicates a rather weak effect of the relaxation of the \textit{4d} MOs, as already discussed at the $GW$ level. Further, the MSD/MAD for TD-CAM-B3LYP and TD-$\omega$B97x with respect to BSE/qs$GW$, are 2 meV/110 meV and -12 meV/129 meV. We conclude from these data that  BSE calculations based on self-consistent ev$GW$@PBE or qs$GW$ calculations come in excellent agreement on the average with each other and with TD-CAM-B3LYP or TD-$\omega$B97x data.

Concerning TD-PBE, the assignment is much more difficult. As a matter of fact, taking again Ag$_4$ as an example, TD-PBE accumulates $\sim$100 excitations below 6~eV, a markedly different behavior as compared to calculations performed with the RSH functionals or with BSE on top of  ev$GW$@PBE or  qs$GW$. In the case of the Ag$_2$ and Ag$_4$ clusters, the TD-PBE onset of absorption is in excellent agreement with other techniques, but for larger clusters the identification even at low energy is more difficult.  As such, the agreement between BSE and TD-PBE is much less satisfactory.  

While the comparison between gas phase calculations performed on the same geometries allows straightforward comparisons, the confrontation between theory and experiment is made difficult by the effect of the rare gas matrix. We provide in Fig.~S4-S5 (SM) a comparison of the calculated gas phase data with the experimental absorption spectra obtained in neon from Ref. \citenum{Lecoultre2011}.  We first focus on the dimer case which comes as an exception since gas phase experiments  are available (see the collection of data in Table~1 Ref.~\citenum{Rabilloud2013}). Further, in the dimer case, the correct identification of the isomers probed experimentally is not an issue. 
BSE on top of self-consistent $GW$ and TDDFT calculations perform very well for the dimer onset of absorption, with a  spread from 2.79~eV (BSE/ev$GW$@PBE) to 2.98~eV (TD-PBE), to be compared to 2.85~eV experimental value. 
At higher energy,  experiments display 4 absorption lines between 4.44~eV and 4.98~eV (gas phase values) matching rather well two doubly-degenerate peaks in the same energy range for the BSE  and the TD-RSH calculations.   As compared to experiment, \cite{degenerate} the MSD/MAD  for these 5 excitations amount to -0.07~eV/0.09~eV for BSE/ev$GW$@PBE and -0.01~eV/0.09~eV for BSE/qs$GW$, an agreement for both methods that is excellent. The smallest deviations at the TD-DFT level is slightly larger, namely -0.20~eV/0.26~eV for TD-CAM-B3LYP.
Concerning TD-PBE, it is difficult to calculate a deviation including states above the onset. 
In particular, TD-PBE only yields one doubly degenerate peak at 4.57~eV. Another doubly degenerate excitation can be found at lower energy (3.67~eV) but with a  nearly vanishing oscillator strength (0.003).

Overall, for the Ag$_n$ (n=2,4,6,8b) clusters,  the MSD/MAD associated with BSE/ev$GW$@PBE   as compared to experiment \cite{degenerate} amount to -0.10~eV/0.11~eV.  The BSE/qs$GW$ results lead to similar results  (-0.10~eV/0.13~eV). This indicates that theoretical values are slightly red-shifted as compared to experiment, keeping in mind that  experiments are performed in matrix for n=4,6,8. In the case of the dimer, experimental results indicate that the matrix induces an overall 0.11~eV blue-shift (see Table~1 Ref.~\citenum{Rabilloud2013}). Such a value is however difficult to adopt for the n=4,6,8 clusters, since the effect of the matrix was shown to depend  on the cluster size, \cite{Schira2018} with red-shifting polarization effects becoming dominant in the large cluster size.  
 QM/QM/PCM calculations \cite{Schira2018} lead to the conclusion that for (n=8) the effect of the matrix is small and the two competing effects tend to cancel out, with a small red-shift for the onset of absorption  but a small blue-shift for the main absorption peaks at $\sim$4~eV. As such, 0.11~eV represents an upper-bound for the (blue-shifting) effect of the matrix on the n=4,6,8 clusters. This  suggests that the agreement between BSE/ev$GW$@PBE or BSE/qs$GW$ with gas phase experiments should fall below a 0.1~eV mean-deviation.  
 TD-CAM-B3LYP offers similar MSD/MAD values of -0.14~eV/0.16~eV as compared to experiment, with the same conclusion that removing the effect of the matrix is expected to bring these deviations below 0.1~eV. 

We finally compare in Fig.~\ref{fig:fig4}(c) various BSE calculations starting either from PBE, PBE0 and PBEh(0.9), focusing on Ag$_4$. The spectra are rather close. This is an illustration of the self-consistent nature of the qs$GW$ or ev$GW$ schemes which fully or partially remove the starting point dependency. BSE/ev$GW$@PBE seems to be in better agreement with the BSE/qs$GW$ data concerning the energy position of the onset of absorption and the structure at $\sim$4~eV,  the typical plasmonic energy range in the limit of larger clusters. The BSE/ev$GW$@PBE0 and BSE/$G_0W_0$@PBEh(0.9) lead to a slight red-shift of the corresponding absorption peaks.  We further consider the  BSE/KS@PBEh(0.9) data, starting with the optimally tuned PBEh(0.9) functional but skipping the $GW$ cycle. The onset of absorption is reasonable, but the peaks structure at higher energy differ  significantly from the BSE/qs$GW$ and BSE/ev$GW$ data. As mentioned above, the problem with the 4\textit{d}-band position at the KS-PBEh(0.9) level can certainly explain this discrepancy. However, a simple non-self-consistent $G_0W_0$@PBEh(0.9) calculation leads to a very reasonable BSE absorption spectrum, an improvement that we associate again with the improved position of the 4\textit{d}-band.

\section{Conclusions}

 Overall, we find that spectra based on standard BSE calculations, i.e. performed in the adiabatic approximation while neglecting further the higher-order ($\partial W / \partial G$) term and vertex corrections in general, agree well with both the experimental results and their best TD-DFT calculation counterparts\cite{Rabilloud2013} based on RSH functionals.   
 However, we find necessary to introduce self-consistency while performing the corresponding $GW$ calculations, especially in the case  of a starting XC functional containing a small amount of exact exchange.
Confirming a previous $GW$/BSE study,\cite{Tiago2009}  we observe that  standard BSE/$G_0W_0$@PBE  fails to produce reasonable absorption spectra, with a collapse of spectral weight at too low energy. As compared to experiment, the BSE/ev$GW$@PBE and BSE/qs$GW$ data are red-shifted by $\sim$0.1~eV (mean-signed-deviation). Considering that 0.1~eV is an upper-bound for the matrix induced blue-shift in the n=4,6,8 cases, the agreement between BSE and experiment is expected to fall below 0.1~eV upon  correcting for  the effect of the rare-gas matrix encapsulating the clusters at the experimental level. 

The present choice of starting with the PBE functional, in this limit of small systems with large gaps and occupied 4\textit{d}-levels, may seem inappropriate. We find however that self-consistency  at the $GW$ level  significantly reduces the impact of the starting functional, bringing BSE/ev$GW$@PBE and BSE/qs$GW$ in excellent agreement with each other and with experiments. 
As such, the relaxation at the qs$GW$ level of the PBE Kohn-Sham molecular orbitals, and the 4\textit{d}-related ones in particular,  has a marginal effect. In fact, the choice of the best starting functional is not an easy question.  In particular, in the larger  size limit,   silver clusters become metallic with vanishing gap and an exponentially decaying screened Coulomb potential. As such, functionals with exact exchange in the long-range may be questioned. However,   short-range exchange should be preserved to stabilize the occupied 4\textit{d}-band.  These considerations need however to be modulated by the fact that TD-DFT calculations with RSH functionals were shown to lead to excellent results in the larger cluster size limit. \cite{Schira2019} Very good results could also be obtained in the large size limit using a local functionals with no long-range exchange but an orbital-dependent onsite $U$ correction for the \textit{d}-levels. \cite{Chaudhary2024}  It is remarkable that  self-consistency at the $GW$ level allows   bypassing the choice of the starting functional,   adapting automatically both the Green's function  and the screened-Coulomb potential to the electronic and dielectric (screening) properties at all-ranges of the systems of interest.

The present study opens the way to explore at the BSE level larger size clusters with emerging plasmonic modes.
It is worthwhile mentioning that low-scaling $GW$  techniques (cubic to below) have been recently developed, 
\cite{Rojas1995,Foerster2011,Neuhauser2014,Vlcek2017,Wilhelm2018,Kim2020,Forster2020,Kutepov2020,Gao2020,Scott2023}   allowing to study systems comprising hundreds of atoms with reasonable computer resources. 
Further, real-time (RT)  BSE calculations have already been implemented. \cite{Attaccalite2011,Umari2021} This allows exploring more efficiently the macroscopic response of larger systems to a  frequency-dependent perturbation, without explicit diagonalization of the  eigenvalue problem expressed in the occupied-virtual product-space. \cite{Casida1995}   This has become one of the major tool to explore at the   RT-TD-DFT level the response properties of large clusters. \cite{Chaudhary2024} 
The development of optimized real-space grids for silver and gold, following Ref.~\citenum{Duchemin2021} where a specific separable-resolution-of-the-identity implementation of cubic-scaling $GW$ was implemented, and the development in the {\sc{beDeft}} code of real-time-BSE techniques, stands as a preliminary pathway to the study of larger clusters. 

\section*{SUPPLEMENTARY MATERIAL}
See the Supplementary Material for (I) the difference between the aQZ and aTZ $GW$ values for the IP, EA and (IP-EA) gap, (II) the full map of the MAD and MSD between methods for the IP, EA and (IP-EA) gap, (III) the analog of Fig.~\ref{fig:fig2} for the Ag$_8$ T$_d$ cluster, (IV) the Ag$_8$ T$_d$ and D$_{2d}$ B3LYP/aDZ anionic structures, (V) the absorption spectra including experimental data from Fig.~1 of Ref.~\citenum{Lecoultre2011}. The peaks used for statistics are indicated by blue/green dots with energy and oscillator strength collected in Table~S2.

\begin{acknowledgments}
The  authors are indebted to Franck Rabilloud for suggesting several references. 
Computational resources were generously provided by  the national HPC facilities under contract GENCI-TGCC A0110910016.
\end{acknowledgments}

\section*{Data Availability Statement}


The data that support the findings of this study are available within the article  and its supplementary material.

 
%



\begin{thebibliography}{127}
\bibitem [{deg()}]{degenerate}%
  \BibitemOpen
  \href@noop {} {}\bibinfo {note} {When comparing theoretical approaches, doubly/triply degenerate excitations only count once in the averages. When comparing to experiments, if a 2-fold/3-fold theoretical excitation matches the energy of 2 very close experimental excitation lines, then they will count twice in the averages. This is the case for the dimer 2nd/3rd bright theoretical excitations, and the Ag$_8$ T$_{d}$ (8a) 1st bright theoretical excitation.}\BibitemShut {Stop}%
\end{thebibliography}
\end{document}